\documentclass{jfm}

\usepackage{graphicx}
\usepackage{natbib}
\usepackage{subfigure}
\usepackage{epstopdf}
\usepackage{mathrsfs} 
\usepackage{rotating} 

\ifCUPmtlplainloaded \else
  \checkfont{eurm10}
  \iffontfound
    \IfFileExists{upmath.sty}
      {\typeout{^^JFound AMS Euler Roman fonts on the system,
                   using the 'upmath' package.^^J}%
       \usepackage{upmath}}
      {\typeout{^^JFound AMS Euler Roman fonts on the system, but you
                   dont seem to have the}%
       \typeout{'upmath' package installed. JFM.cls can take advantage
                 of these fonts,^^Jif you use 'upmath' package.^^J}%
      }
  \else
  \fi
\fi

\ifCUPmtlplainloaded \else
  \checkfont{msam10}
  \iffontfound
    \IfFileExists{amssymb.sty}
      {\typeout{^^JFound AMS Symbol fonts on the system, using the
                'amssymb' package.^^J}%
       \usepackage{amssymb}%
       \let\le=\leqslant  
         
      }{}
  \fi
\fi

\ifCUPmtlplainloaded \else
  \IfFileExists{amsbsy.sty}
    {\typeout{^^JFound the 'amsbsy' package on the system, using it.^^J}%
     \usepackage{amsbsy}}
    {}
\fi

\newsavebox{\astrutbox}
\sbox{\astrutbox}{\rule[-5pt]{0pt}{20pt}}

\DeclareMathAlphabet{\mathpzc}{OT1}{pzc}{m}{it}

\title[Dynamics of elastocapillary rise]{Dynamics of elastocapillary rise}

\author[C. Duprat, J. M. Aristoff and H. A. Stone]%
{C\ls A\ls M\ls I\ls L\ls L\ls E\ns D\ls U\ls P\ls R\ls A\ls T,\ns
J\ls E\ls F\ls F\ls R\ls E\ls Y\ns M.\ns A\ls R\ls I\ls S\ls T\ls O\ls F\ls F\ls \break
\and H\ls O\ls W\ls A\ls R\ls D\ns A.\ns S\ls T\ls O\ls N\ls E}

\affiliation{Department of Mechanical and Aerospace Engineering, Princeton University,\\ Princeton, NJ 08544, USA}

\begin{document}

\maketitle

\begin{abstract}
We present the results of a combined experimental and theoretical investigation of the surface-tension-driven coalescence of flexible structures. Specifically, we consider the dynamics of the rise of a wetting liquid between flexible sheets that are clamped at their upper ends. As the elasticity of the sheets is progressively increased, we observe a systematic deviation from the classical diffusive-like behaviour: the time to reach equilibrium increases dramatically and the departure from classical rise occurs sooner, trends that we elucidate via scaling analyses. Three distinct temporal regimes are identified and subsequently explored by developing a theoretical model based on lubrication theory and the linear theory of plates. The resulting free-boundary problem is solved numerically and good agreement is obtained with experiments.
\end{abstract}


\section{Introduction}
Surface tension plays a role in an enormous variety of industrial problems and natural phenomena: the string of dew drops on a spider web, the spontaneous imbibition of water into soil, the stability of water-walking arthropods, the wet lithography of micro-electro-mechanical systems, the rise of sap in trees, and the wet adhesion of insects on plants, to name just a few. It is clear that the elasticity of the substrate is relevant to understand some of these phenomena, including the clumping of the flexible bristles of tarsi of insects when tarsal oil is released~\cite[]{Eis00} or the adhesive failure (stiction) of micro-cantilevers bent by capillary forces at the evaporating menisci~\cite[]{Mas93t}. Although static problems involving the interplay between capillarity, elasticity, and possibly other forces are well studied, there are very few studies of the dynamics of elastocapillarity, which is the subject of this paper.

Early studies of elastocapillarity dealt with the deformation of a solid in contact with a liquid drop. From the viewpoint of mechanics, when a liquid drop adheres to a substrate, there must be a balance of the forces acting at the contact line. The familiar Young's equation can be interpreted as a balance of the tangential forces. Some authors~\cite[]{Les61,For84} suggested that the normal component is balanced by a force exerted by the solid, which deforms upon contact. This deformation is generally negligible on a rigid substrate, but can lead to the formation of a ridge~\cite[]{And79,Per08} or wrinkles~\cite[]{Hua07} on a soft substrate. Motivated by the recent interest in soft materials in engineering, nano-fabrication, and biomechanics, several authors have addressed the question of how slender objects deform under capillary forces~\cite[]{Bic04,Kim06,Py07,Pok09}, mostly considering equilibrium configurations. 

Nevertheless, some elastocapillary effects are essentially dynamic (e.g. sap flow, spore release), and it is of fundamental and practical importance to estimate the time to reach equilibrium in such systems, as well as to understand their transient dynamics. Recent examples of such elastocapillary dynamics include studies of capillary-driven flow in deformable porous media~\cite[]{Sid09} and in flexible channels where gravitational effects could be neglected~\cite[]{Hon07, Ari10}. Here, we consider a model system to study elastocapillary dynamics with gravity, namely capillary rise between flexible sheets that are clamped at their upper ends. In doing so, we extend the work of \cite{Bic04} and \cite{Kim06} by considering the \textit{dynamics} of elastocapillary rise. 
The experimental set-up and results are described in $\S$\ref{sec:exp}. We begin with statics, identify three different regimes, and then address the transient behaviour. We show how the evolution of the position of the meniscus departs from that of classical capillary rise (i.e. rise between rigid boundaries) and exhibits an unusual behaviour: the meniscus slows down, reaches a plateau, then reaccelerates before finally stopping. In $\S$\ref{sec:theory}, we formulate a free-boundary problem to model the dynamics, and in $\S$\ref{sec:solution} we describe the solution technique and present the numerical solutions. The transient dynamics is analysed in $\S$\ref{sec:discussion} in light of experimental and numerical results.

\section{Experiments} \label{sec:exp}
\subsection{Experimental setup and dimensionless numbers} \label{sec:setup}
\begin{figure}
\begin{center}
\includegraphics[width=3.7 cm]{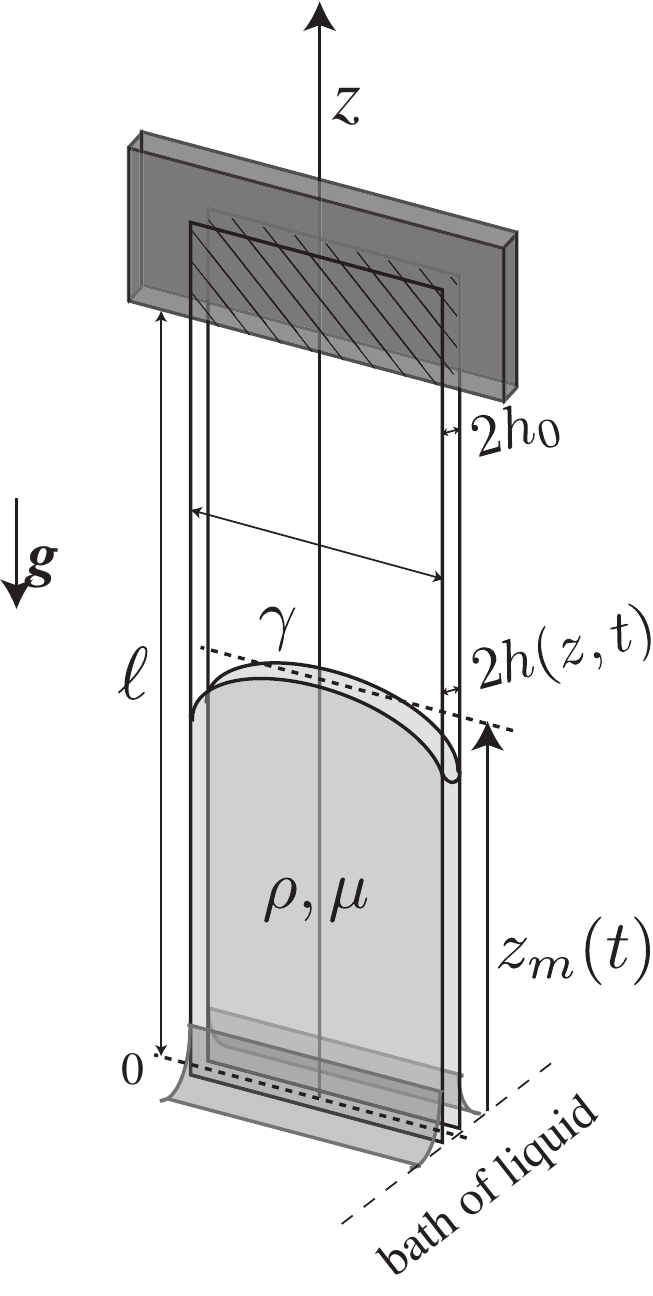}
\caption{Sketch of the experimental setup. The top hatched region corresponds to the clamped edge of the glass sheets. The shape of the meniscus is indicative of a typical experimental observation. We track the position of the apex of the meniscus.}
\label{fig.1}
\end{center}
\end{figure}

In this section we describe our experimental setup to investigate the dynamics of capillary rise between flexible sheets (Fig.~\ref{fig.1}). Two vertical glass sheets (length $\ell$, thickness $b=160~\mu$m, width $w=5$ mm and bending stiffness per unit width $B=(2.2~\pm~0.2)~\times~10^{-2}$ N$\cdot$m) are initially parallel. The sheets are clamped at their upper ends, with a separation distance $2h_0$, and free at their lower ends. At time $t=0$, the lower ends of the sheets are brought into contact with a bath of silicon oil (viscosities and densities $\mu=$ 0.096 Pa$\cdot$s , $\rho=$ 963 kg$\cdot$m$^{-3}$ (V100), $\mu=$ 0.048 Pa$\cdot$s, $\rho=$ 960 kg$\cdot$m$^{-3}$ (V50), $\mu=$ 0.0093 Pa$\cdot$s, $\rho=$ 930 kg$\cdot$m$^{-3}$ (V10), and  surface tension $\gamma=0.021$ N$\cdot$m$^{-1}$) that perfectly wets the glass (i.e. the  equilibrium contact angle $\theta_e=0$). The deflection of the sheets and the position of the meniscus as it rises between the sheets are recorded from the side with a digital camera and tracked using customised image-analysis software in MATLAB\textsuperscript{\textregistered}. 

We take the length of the sheet $\ell$ and half the initial gap $h_0$ as, respectively, the characteristic lengths for the meniscus position $z_m(t)$ and the deflection of the sheets from the vertical $h(z,t)$. We can then estimate  the capillary energy $E_c\sim w \ell \gamma$, the gravitational energy $E_g\sim \rho g w h_0 \ell^2$ and the bending energy $E_e\sim B w h_0^ 2/\ell^3$ for our system, where $g$ is the gravitational acceleration and here we use $\sim$ to indicate order-of-magnitude scaling estimates for which prefactors are typically order one. Balancing $E_c$ with $E_g$, $E_c$ with $E_e$, and $E_g$ with $E_e$ yields three characteristic lengths:
\begin{equation}
\ell_{cg}= \frac{\gamma}{\rho g h_0}, \qquad \ell_{ec}= \left( \frac{Bh_0^2}{\gamma} \right)^{1/4},~\mathrm{and}~~ ~\ell_{eg}= \left( \frac{Bh_0}{\rho g} \right)^{1/5}.
\end{equation}
The capillary-gravity length $\ell_{cg}$ is the classical equilibrium height for the capillary rise between two rigid plates. The elastocapillary length $\ell_{ec}$ corresponds to the minimal length above which capillary forces can bring the sheets together and has been identified previously in studies of the failure of micro-mechanical structures upon drying \cite[]{Mas93t, Boe99} or hair clumping \cite[]{Bic04}. The scale $\ell_{eg}$ is an elastogravity length. Our system is thus described by two dimensionless numbers,
\begin{equation}
\mathcal{B}=\frac{\ell}{\ell_{cg}}~\mathrm{and}~ \mathscr{E}=\left(\frac{\ell}{\ell_{ec}}\right)^4,
\end{equation}
where $\mathcal{B}$ is the Bond number and $\mathscr{E}$ is the elastocapillary number. Note that we may also define an elastogravity number $\mathcal{G}=\left(\ell/\ell_{eg}\right)^5=\mathcal{B} \mathscr{E}$. 
By changing the spacing $0.18~\mathrm{mm}\leqslant h_0 \leqslant 0.39~\mathrm{mm}$, we varied the elastocapillary length $13.4~\mathrm{mm}\leqslant \ell_{ec}\leqslant 20~\mathrm{mm}$. The elastocapillary number $\mathscr{E}$ is varied within a wide range ($0.18\leqslant \mathscr{E} \leqslant 192$) by changing the length of the sheets $10~\mathrm{mm} \leqslant \ell \leqslant 50~\mathrm{mm}$. The value of the Bond number $\mathcal{B}$ is also affected by the changes in $\ell$, but remains of order one ($1\leqslant \mathcal{B} \leqslant 7$) for all experiments. We note that for $\mathcal{B} \leqslant 1$, the meniscus will fill the entire gap between the sheets (i.e. $\ell_{cg}\geqslant\ell$).
\subsection{Equilibrium configuration}
\begin{figure}
\begin{center}
\includegraphics[width=0.95\textwidth]{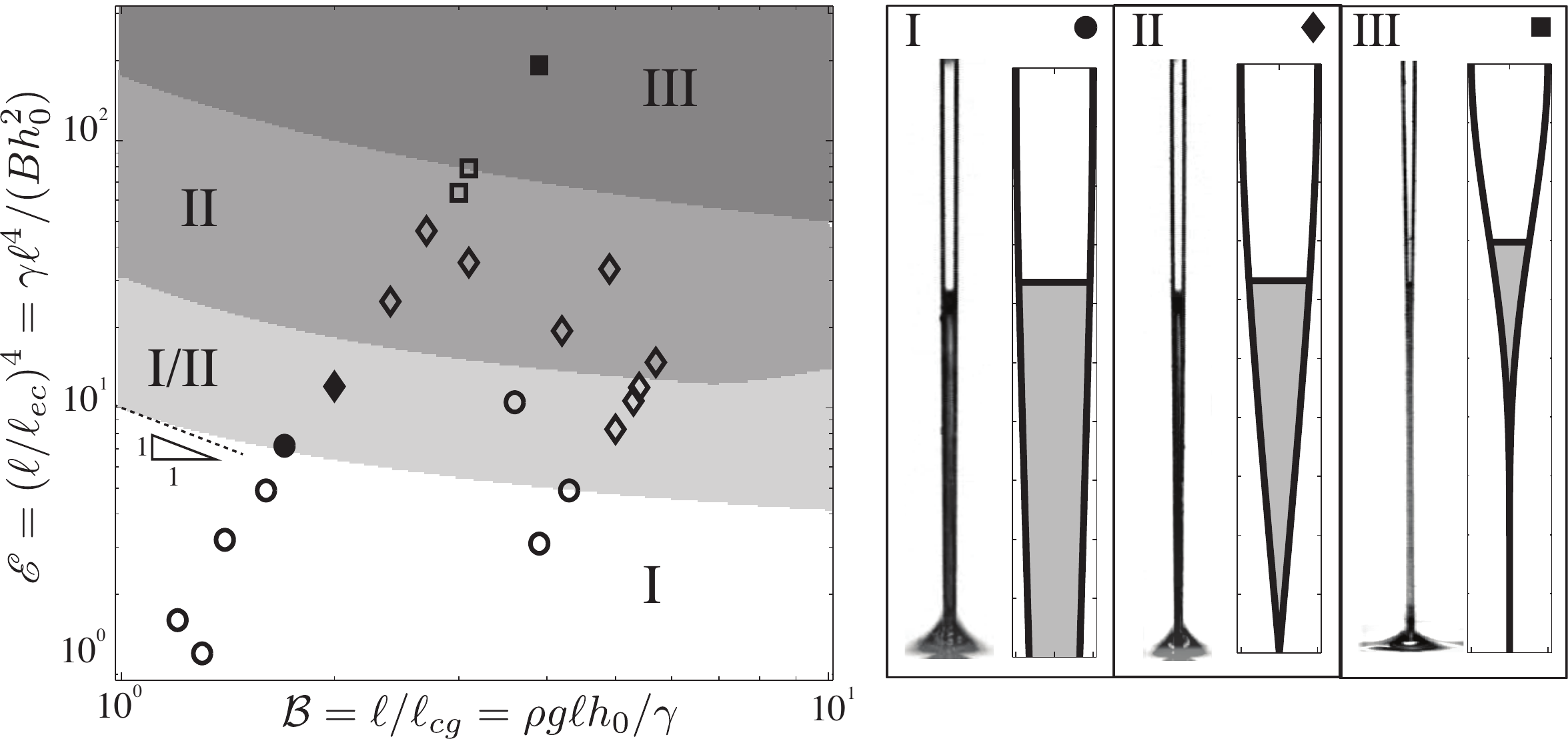}
\caption{Equilibrium configurations. Regime diagram of elastocapillary rise in the parameter space $(\mathscr{E}, \mathcal{B})$ obtained by studying the static problem experimentally (symbols) and numerically (shaded regions). Three regimes, labeled I ($\circ$), II ($\diamond$) and III ($\square$) are identified and illustrated with experimental photographs and numerically deduced profiles  for (I, \Large{$\bullet$}\small) $\mathscr{E}=7.2,~\mathcal{B}=1.7$, (II, $\blacklozenge$) $\mathscr{E}=12,~\mathcal{B}=1.9$ and (III, $\blacksquare$) $\mathscr{E}=192,~\mathcal{B}=3.9$. In the region labeled I/II, we predict numerically the coexistence of two regimes (I and II), in agreement with our experimental observations.}
\label{fig.2}
\end{center}
\end{figure}

When the free ends of the sheets are brought into contact with the bath, the liquid spontaneously rises up, and the induced pressure distribution leads to an inward deflection of the sheets until the meniscus  stops at a finite height. This equilibrium state depends on the combined effects of surface tension, gravity, and elasticity. Depending on the two dimensionless parameters ($\mathscr{E}$, $\mathcal{B}$), three different configurations are observed, as depicted in Fig.~\ref{fig.2}. For $\mathscr{E} \lesssim 10$, the sheets are slightly deflected while the lower ends remain open (Regime I). For $\mathscr{E} \gtrsim 100$, the sheets deflect and coalesce over a distance comparable to their length (Regime III). Nevertheless, the sheets do not make dry contact; a thin liquid film remains between them. The statics of these two regimes have been investigated previously \cite[]{Bic04,Kim06}. In addition, we have identified an intermediate regime (Regime II) wherein the sheets touch at the tip, but there is a nonzero angle between them. 

The equilibrium shape of the sheets is found by solving numerically the nonlinear free-boundary problem developed by~\cite{Kim06}, with a different boundary condition at the lower ends of the sheets necessary to find an equilibrium shape for intermediate $\mathscr{E}$, which corresponds to Regime II (see $\S$\ref{sec:theory} and Table~\ref{problemsummary}). We compiled the permissible equilibrium configuration(s) for 120,000 different pairings of $\mathscr{E}$ and $\mathcal{B}$, and present the results in Fig.~\ref{fig.2} (shaded regions). Good agreement with our experimental observations is observed. We note that for $10\lesssim \mathscr{E} \lesssim 30$, solutions for regimes I and II coexist.

\cite{Kim06} performed a scaling analysis to determine the onset of large deflection of the sheets, which can be interpreted as the upper boundary of Regime I. To do so they balanced the torque (per unit width) exerted on the sheets, $B h_0/\ell^2$, with the capillary-gravity torque, $\rho g \ell_{cg}^2 (\ell-\ell_{cg}/2)$. They considered the limit $\ell \gg \ell_{cg}$, (i.e. $\mathcal{B}\gg 1$), wherein the capillary-gravity torque becomes $\rho g \ell_{cg}^2 \ell$, and thus they obtained a boundary that is given by $\mathscr{E}\sim \mathcal{B}$. Our numerical results suggest that this scaling is indeed valid for large Bond numbers (not shown). However, our experiments and numerically predicted equilibrium shapes show a different behaviour for $\mathcal{B}\lesssim 10$, which can be understood by considering the limit $\ell/\ell_{cg} \rightarrow 1$ (i.e. $\mathcal{B} \rightarrow 1$), wherein the capillary-gravity torque becomes $\rho g \ell^3$. Hence, we find a boundary that is given by $\mathscr{E} \sim \mathcal{B}^{-1}$, in agreement with our numerical solutions shown in Fig.~\ref{fig.2}. 

\subsection{Dynamics}
\begin{figure}
\begin{center}
\includegraphics[width=0.9\textwidth]{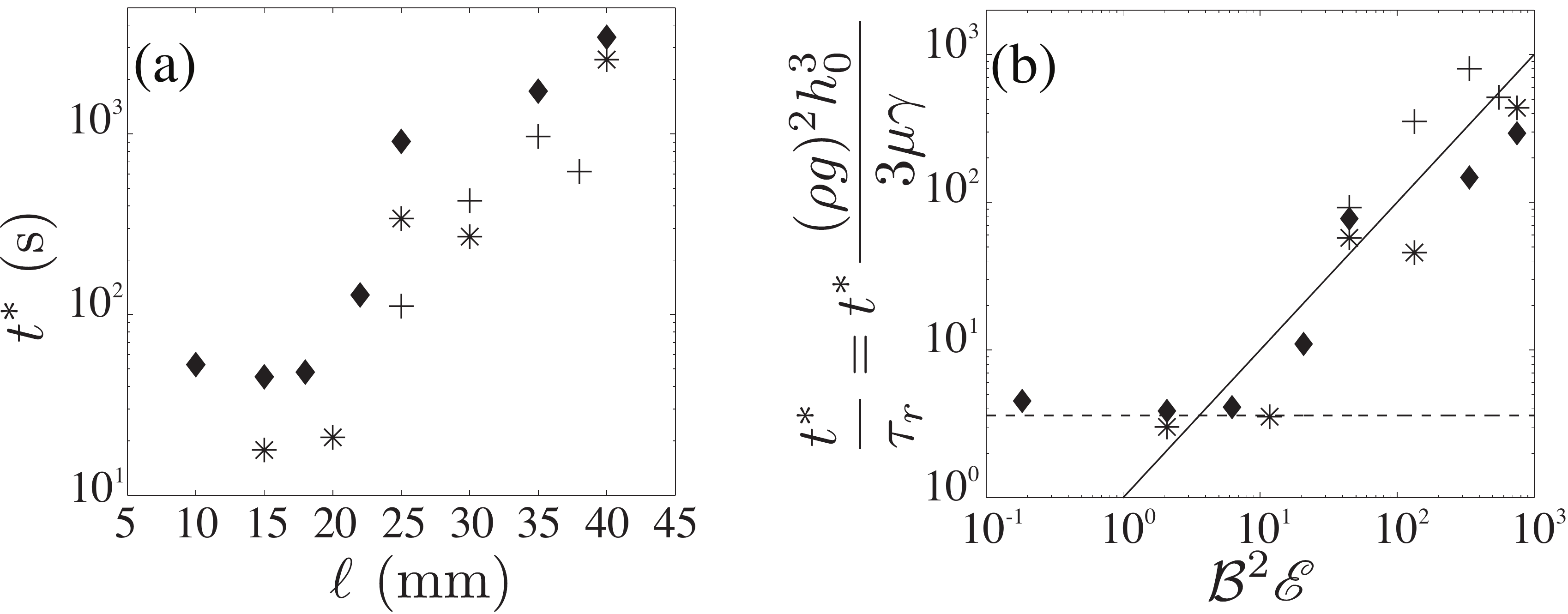}
\caption{Time to reach equilibrium. (a) Time $t^*$ in seconds (s) required to reach 99$\%$ of the equilibrium height versus the length of the sheet  $\ell$ in millimeters (mm). Data points correspond to experiments with a constant initial gap $h_0=0.18$ mm and with silicone oil of viscosity V100 ($\blacklozenge$), V50 ($\ast$) and V10 ($+$). (b) Time $t^*$ scaled by the characteristic time to reach equilibrium in the classical rigid case, $\tau_r$, versus $\mathcal{B}^2\mathscr{E}$. The dotted line corresponds to $t^*=3.6 \tau_r$ and the solid line to $t^*/\tau_{r}=\mathcal{B}^2\mathscr{E}$, i.e. $t^*=\tau_{ve}$.}
\label{fig.2b}
\end{center}
\end{figure}

We now focus on dynamics, in particular the time to reach these equilibrium configurations. We measured the time $t^*$ taken by the meniscus to reach 99$\%$ of its final height. Results obtained for $h_0=0.18$ mm and three different viscosities are presented in Fig.~\ref{fig.2b}(a), where we varied the length of the sheet $\ell$ and hence the elastocapillary number $\mathscr{E}$. For short sheets, $t^*$ is roughly constant and on the order of a minute. The sheets deflect slightly and the final state is close to the one obtained for rigid boundaries (Regime I). For larger sheets, the deflection becomes important (Regimes II and III) and $t^*$ increases rapidly with increasing $\ell$.

To estimate the time to reach equilibrium, we assume that the speed of the meniscus $u$ is given by that for a Poiseuille flow $u=\Delta p h_0^2/(3\mu\ell)$. Provided the deflection of the sheets is small, the pressure difference across the meniscus may be approximated by $\Delta p=\gamma/h_0$, and the characteristic length is the capillary-gravity length $\ell_{cg}$. The time to reach equilibrium $\tau_r$ is thus given by a balance between capillary, viscosity, and gravitational forces, leading to $\tau_r=\frac{3\mu \gamma}{h_0^3(\rho g)^{2}}$ \cite[]{Que97}, which is independent of the length of the sheets $\ell$.

When the deflection of the sheets is appreciable, the two dominant effects are viscosity and elasticity. The pressure difference $\Delta p$ is set by the typical deflection of the sheets $h_0$, i.e. $Bh_0/\ell^4=\Delta p$. Therefore, we can define a characteristic visco-elastic time $\tau_{ve}=\frac{\ell}{u}=\frac{3\mu\ell^6}{Bh_0^3}=\mathcal{B}^2\mathscr{E} \tau_r$, which rapidly increases with increasing $\ell$. 

The time to reach equilibrium, rescaled as $t^*/\tau_r$, is presented in Fig.~\ref{fig.2b}(b) versus $\mathcal{B}^2\mathscr{E}$. The data follow two different behaviours. For $\mathcal{B}^2\mathscr{E}\lesssim 10$, we recover the classical result for capillary rise between rigid sheets: $t^*\simeq \tau_r$, indicating that elastic effects are negligible. Conversely, for $\mathcal{B}^2\mathscr{E}\gtrsim 10$ the time to reach equilibrium scales with the visco-elastic time, i.e. $t^*/\tau_r=\mathcal{B}^2\mathscr{E}$, indicating that elastic effects are dominant.

\begin{sidewaysfigure}
\centering
\vspace{5.15in}
\includegraphics[width=0.89\textwidth]{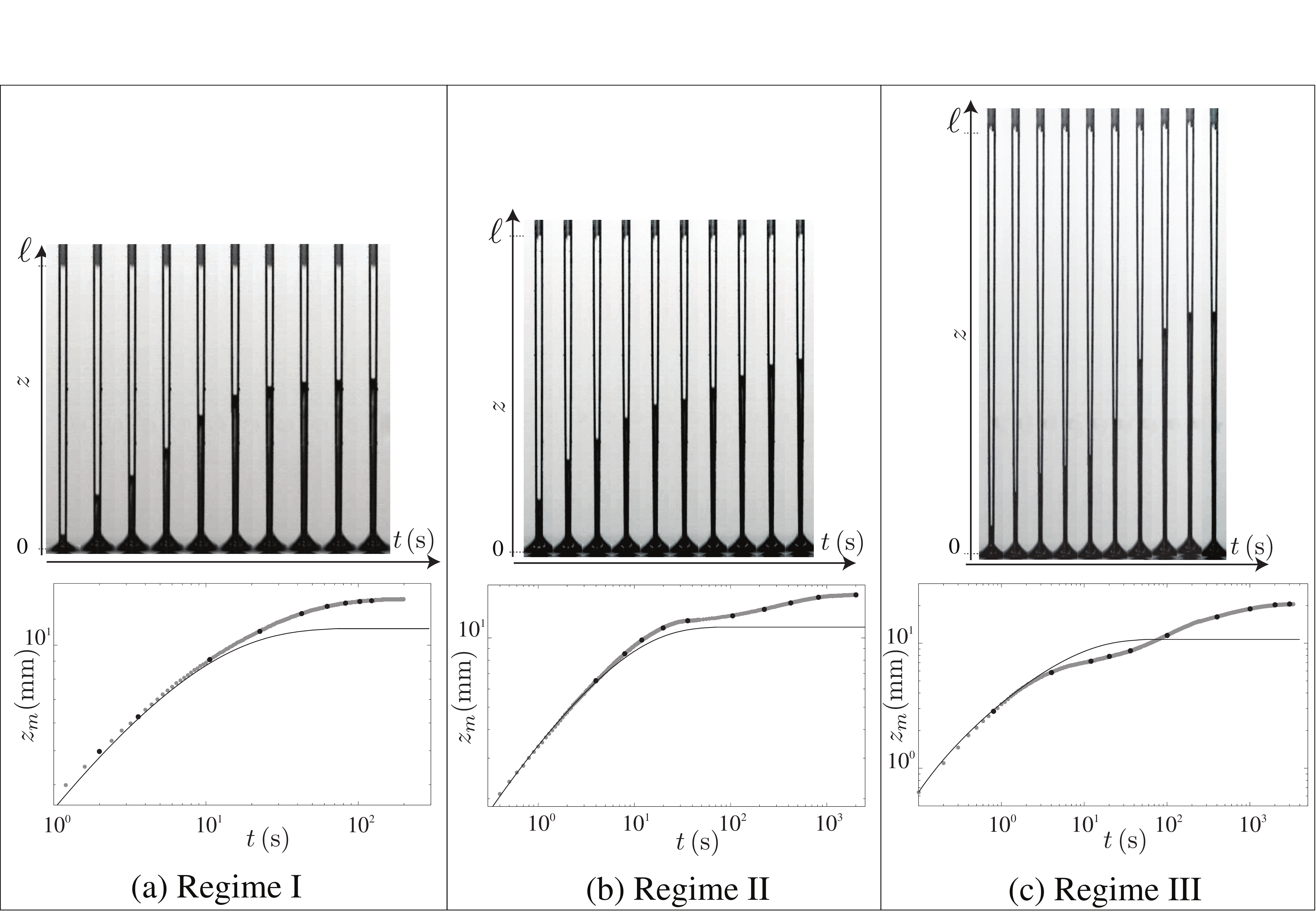}
\hspace{1in}\parbox{6.5in}{
\caption{Image sequences of elastocapillary rise and time evolution of the position of the meniscus $z_m$ for $h_0=0.18$ mm, silicone oil V100 ($\ell_{ec}=13.4$ mm) and (a) Regime I, $\ell=22$ mm, $\mathscr{E}=6.8$, $\mathcal{B}=1.8$, (b) Regime II , $\ell=25$ mm, $\mathscr{E}=11.4$, $\mathcal{B}=2$  and (c) Regime III, $\ell=35$, $\mathscr{E}=43.9$, $\mathcal{B}=2.8$. The times corresponding to each image are denoted by the ($\bullet$) symbols on the grey curves. For comparison, the solid curves denote the corresponding meniscus position for capillary rise between rigid sheets, given by (\ref{speedandposition}).} \label{fig.4}}
\end{sidewaysfigure}

Next, we examine the evolution of the shape of the sheets and the position of the meniscus for three experiments representative of Regimes I, II and III (Fig.~\ref{fig.4}). The dynamics of imbibition between flexible sheets is compared to the classical case of rise between rigid sheets (black lines in Fig.~\ref{fig.4}), wherein the meniscus follows a diffusive-like behaviour at early times, i.e. $z_m\propto t^{1/2}$ \cite[]{Bell1906}, and is then slowed by gravity until eventually reaching its equilibrium height $\ell_{cg}$.
In our experiments (grey points in Fig.~\ref{fig.4}), the meniscus follows this classical behaviour at early times, as  the sheets do not appreciably deflect. At later times, the sheets begin to bend and the meniscus position deviates from that of classical capillary rise. For small $\mathscr{E}$ (Fig.~\ref{fig.4}(a)), the sheets are slightly deflected. The meniscus slows down and stops at a higher position than the classical height. For intermediate $\mathscr{E}$ (Fig.~\ref{fig.4}(b)), the ends of the sheets touch during imbibition. The meniscus slows down, reaches a plateau, then surprisingly, accelerates until finally stopping. For large $\mathscr{E}$ (Fig.~\ref{fig.4}(c)), the sheets touch soon after imbibition begins, when the meniscus is quite close to the bath and gravity has not yet contributed to its deceleration. At intermediate times, the position of the meniscus dips below that of classical rise and the meniscus slows down considerably. After roughly one minute has elapsed, the sheets begin to coalesce and the meniscus accelerates until eventually reaching its equilibrium height.

\begin{figure}
\begin{center}
\includegraphics[width=0.8\textwidth]{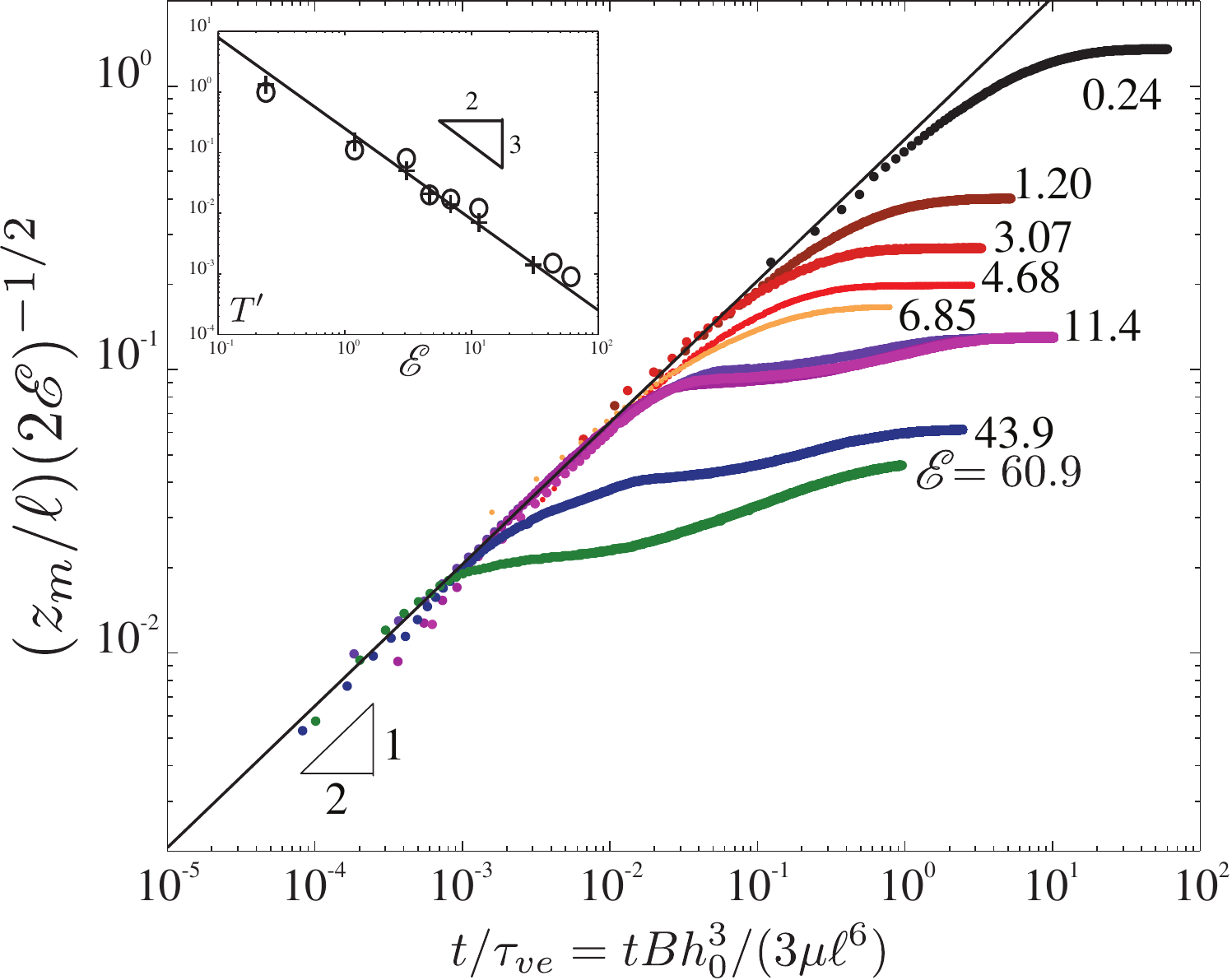}
\caption{Rescaled position of the meniscus $(z_m/\ell) (2\mathscr{E})^{-1/2}$ as a function of time $T=t/\tau_{ve}=t~Bh_0^3/(3\mu\ell^6)$. The solid line corresponds to a diffusive-like behaviour $Z_m \propto \sqrt{\mathscr{E} T}$.
Inset: evolution of the cross-over time $T'$ at which the meniscus position deviates from the classical prediction as a function of $\mathscr{E}$, for experiments ($\circ$) and numerics ($+$). Numerical solutions are described in \S\ref{sec:solution}. The solid line represents the predicted scaling $T'\propto ~\mathscr{E}^{-3/2}$.}
\label{fig.5a}
\end{center}
\end{figure}

To obtain a better understanding of the dynamics of elastocapillary rise, we compare the motion of the meniscus for several experiments as the elastocapillary number $\mathscr{E}$ is progressively increased. Since the meniscus position at early times is given by $z_m\propto t^{1/2}$, or equivalently, in dimensionless terms, $Z_m=z_m/\ell\propto \mathscr{E}^{1/2}(t/\tau_{ve})^{1/2}$, we use the characteristic length and time scales $\ell \mathscr{E}^{1/2}$ and $\tau_{ve}$, respectively, to rescale the experimental data. The evolution of the dimensionless position of the meniscus $(z_m/\ell) (2\mathscr{E})^{-1/2}$ with time $T=t/\tau_{ve}$ is presented in Fig.~\ref{fig.5a}. 
The family of curves obtained for various $\mathscr{E}$ values collapse onto the same line at early times, corresponding to a diffusive-like behaviour of the advancing meniscus, i.e. $Z_m \propto \sqrt{\mathscr{E} T}$. Here, we have used $\propto$ to distinguish order-of-magnitude results from results indicating proportionality.
We have also verified that the fluid viscosity enters the dynamics only through $\tau_{ve}$, as shown by the collapse, in Fig.~\ref{fig.5a}, of the curves for three different viscosities (V10, V50, and V100) at $\mathscr{E}=11.4$.

The position of the meniscus deviates from $T^{1/2}$ at a time $T'$. Increasing $\mathscr{E}$ decreases this cross-over time $T'$, as the sheets bend more easily. Neglecting the deceleration due to gravity, we expect this deviation to occur when the deflection of the sheets due to capillary forces becomes important, that is for $z_m\sim \ell_{ec}$, or in dimensionless terms, $Z_m \sim \mathscr{E}^{-1/4}$. Before reaching this value, the meniscus evolves accordingly to $T\propto Z_m^2/\mathscr{E}$. Hence, we expect a cross-over time of $T'\propto\mathscr{E}^{-3/2}$. The dependence of $T'$ on $\mathscr{E}$ is given in the inset of Fig.~\ref{fig.5a}. The predicted scaling is indeed recovered. For times greater than the cross-over time, the speed of the meniscus slows down dramatically. As the elastocapillary number $\mathscr{E}$ is increased, the motion of the meniscus becomes increasingly complex. In particular, we observe a plateau region followed by an acceleration of the meniscus. In order to understand this behaviour quantitatively, we now formulate a theoretical model for the dynamics of elastocapillary rise. 
\section{Theoretical formulation} \label{sec:theory}
In this section we formulate a free-boundary problem that describes the evolution of the  shape of the sheets and the rise of liquid between them. We first consider the deflection of the initially vertical sheets. Owing to the symmetric configuration, we need only consider one of the sheets. Provided that the initial gap, $2h_0$ (set by the clamping distance), is much less than the length of the sheet, $\ell$, the slope of the sheet, $h_0/\ell$, is small. We assume that the sheet is sufficiently thin, $b\ll w\ll \ell$, so that we may use the linear theory of beams to describe the deflection, $h(z,t)$, from the vertical. Hence,
\begin{equation}
Bh_{zzzz}+m h_{tt}=p(z,t), \label{beameqn}
\end{equation}
where subscripts denote partial derivatives, $B=Eb^3/12(1-\nu^2)$ is the bending stiffness per unit width, $E$ the Young's modulus, $\nu$ the Poisson ratio, $m$ the mass per unit length per unit width, and $p(z,t)$ the force per unit area (pressure) acting on the sheet. The second term in (\ref{beameqn}) arises from the inertia of the beam, and may be neglected if the time scale of the flow, $\tau_{ve}$, is much longer than the reaction time of the beam to an applied load, $\tau=\sqrt{\frac{m \ell^4}{B}}$, obtained by balancing the characteristic bending and inertial stresses. We use this quasi-static description of the sheet wherein $\tau_{ve} \gg \tau$. Hence, in the following,
\begin{equation}
Bh_{zzzz}=p(z,t). \label{beameqnstatic}
\end{equation}
We note that the weight of the sheet gives rise to a stretching strain of order $mg\ell/Eb$. Comparing this to a typical bending strain $Bh_0/\ell^2$ leads to a maximum length $\ell_s\sim\left(Eb^2h_0/mg\right)^{1/3}$, below which stretching effects can be ignored. In our experimental study, we have $\ell \ll \ell_s\simeq 40$ cm.

We now consider the incompressible, viscous flow of a perfectly wetting liquid $(\theta_e=0)$ that enters the region between the sheets owing to a reduction in pressure at the meniscus relative to the bath. Our results are extendable to partially wetting liquids $(0<\theta_e<\pi/2)$. In what follows, we take the limit $h_0 \ll w$, relevant to our experiments, so that edge effects may be safely neglected. Hence, the pressure drop is given by $\gamma\kappa$, where $\kappa = \cos\theta_d/h(z_m,t)$ is the curvature at the meniscus $z=z_m(t)$ and $\theta_d$ is the apparent dynamic contact angle. Dynamical features of the moving contact line are treated by using a semi-empirical relationship for the case of total wetting, in which the dynamic contact angle, $\theta_d$, is defined as 
\begin{equation}
\theta_d=\left(\frac{\Gamma\mu}{\gamma}\frac{dz}{dt}\right)^{1/3}, \label{tanner}
\end{equation}
as first measured by~\cite{Hof75} and later established by~\cite{Tan79}. This relationship accounts for viscous dissipation at the contact line and leads to a reduced driving pressure during imbibition \cite[]{Guy01}. The parameter $\Gamma=6 \ln{(x_M/x_m)}$ arises from evaluating the viscous dissipation near the contact line, where it is necessary to introduce the upper $x_M$ and lower $x_m$ cut-off lengths to prevent the divergence of the integral measuring the total viscous dissipation. No exact solution for the microscopic description is currently available, and thus we treat $\Gamma$ as an adjustable parameter. In order to determine its value for our experimental study, we performed a series of control experiments where we examined the imbibition of silicone oil between rigid glass boundaries, both in planar and axisymmetric configurations, and with and without gravity. The value of $\Gamma$ which best describes the results of these control experiments is $\Gamma=60$, and so we take it to have this value in our numerical simulations of elastocapillary rise. This value of $\Gamma$ is comparable with those found in the literature. Finally, we note that our numerical results are only slightly dependent on the value of $\Gamma$; simulations with $50<\Gamma<70$ give similar results.

 A one-dimensional approximation to the viscously dominated flow is given by Darcy's law,
\begin{equation}
\frac{\mu u}{k}=-\frac{\partial p}{\partial z}-\rho g, \label{darcy}
\end{equation}
where $u$ is the cross-sectionally averaged liquid velocity, $p$ the local pressure and $z$ the axial coordinate. The permeability, $k=h^2/3$, where $2h$ is the total distance between the sheets, is based on the lubrication approximation \cite[]{Guy01} which requires $|h_z| \ll 1$, and is consistent with our description of the sheets.

Differentiating (\ref{beameqnstatic}) once in $z$ and combining the result with (\ref{darcy}) gives an expression for the liquid velocity,
\begin{equation}
u(z,t)=-\frac{h^2}{3\mu}\left(B h_{zzzzz}+\rho g\right), \label{velocity}
\end{equation} 
which may be evaluated at $z=z_m(t)$ to find an expression for the speed of the meniscus.
Next, combining (\ref{velocity}) with the statement of mass conservation, $h_t+(hu)_z=0$, yields a nonlinear evolution equation for the deflection of the sheets in the liquid-filled region:
\begin{equation}
h_t=\frac{h^2}{3\mu}\left(Bhh_{zzzzzz}+3Bh_z h_{zzzzz}+3\rho g h_z\right).
 \label{evolution}
\end{equation}
Let $\tilde{h}(z,t)$ denote the deflection of the sheets in the liquid-free region. Using (\ref{beameqnstatic}) with $p(z,t)=0$, we have 
\begin{equation}
B\tilde h_{zzzz}=0,
\end{equation}
which may be integrated analytically. Equation (\ref{evolution}) is similar to the equation studied by~\cite{Hosoi04}, who considered the dynamics of an elastic sheet that was clamped at one end and lubricated from below by a single fluid. Our study is distinguished by the influence of surface tension and the presence of two fluids whose boundary advances according to (\ref{velocity}) and depends on the local shape of the sheet.

The evolution equation and boundary conditions may be rendered dimensionless using the characteristic length and time scales presented in \S\ref{sec:setup}. Letting $H=h/h_0$, $Z=z/\ell$, $Z_m=z_m/\ell$, and $T=t/\tau_{ve}$, transforms (\ref{velocity}), evaluated at the meniscus, to 
\begin{equation}
\frac{dZ_m}{dT}=-H^2\left(H_{ZZZZZ}+\mathscr{E}\mathcal{B} \right)\Big |_{Z=Z_m} \label{velocityND}
\end{equation}
and (\ref{evolution}) to
\begin{equation}
H_T=H^3H_{ZZZZZZ}+3H^2H_{Z} H_{ZZZZZ}+3\mathscr{E}\mathcal{B} H^2 H_{Z}. \label{evolutionND}
\end{equation}
Two initial conditions and six boundary conditions are necessary to solve the free-boundary problem specified by (\ref{velocityND}) and (\ref{evolutionND}). The first two boundary conditions are $H_{ZZ}(0,T)=0$, and $H_{ZZZZ}(0,T)=0$. These correspond, respectively, to zero moment and zero pressure (relative to the atmosphere) at the lower end of the sheets. If the free ends of the sheets remain open (Regime I), we have the condition $H_{ZZZ}(0,T)=0$, which corresponds to zero force at the lower end of the sheets. If the sheets are in contact (Regime II), they exert an equal but opposite force on each other, and the force-free condition $H_{ZZZ}(0,T)=0$ must be replaced by $H(0,T)=0$.  If the sheets are in contact over a finite length $0<Z\le Z_c(T)$, where $Z_c=z_c/\ell$ and $z_c$ is the highest point at which the sheets contact, as in Regime III, we have an additional unknown, $Z_c$, and an additional condition $H_Z(Z_c,T)=0$, that enforces a tangentially smooth contact between the sheets.
At the meniscus, the jump in pressure given by (\ref{beameqnstatic}) is equal to the Laplace pressure $\gamma \kappa$, which can be written in dimensionless form as $H_{ZZZZ}(Z_m,T)-\tilde{H}_{ZZZZ}(Z_m,T)=-\mathscr{E}\cos \theta_d/H(Z_m)$.

The remaining two boundary conditions for $H(Z,T)$ are obtained by expressing the matching conditions at the meniscus between the liquid-filled and the liquid-free regions in terms of the boundary conditions at the top end of the sheet (fixed and clamped, i.e. $\tilde{H}(1,T)=1$ and $\tilde{H}_Z(1,T)=0$). Combining the conditions of equal deflection $\tilde{H}(Z_m,T)=H(Z_m,T)$, equal slope $\tilde{H}_Z(Z_m,T)=H_Z(Z_m,T)$, equal moment $\tilde{H}_{ZZ}(Z_m,T)=H_{ZZ}(Z_m,T)$ and  equal shear force (since the liquid is perfectly wetting) $\tilde{H}_{ZZZ}(Z_m,T)=H_{ZZZ}(Z_m,T)$, yields 
\begin{equation}
H(Z_m,T)=1-\frac{1}{3}(H_{ZZZ}(Z_m,T))(Z_m-1)^3+\frac{1}{2}H_{ZZ}(Z_m,T)(Z_m-1)^2
\end{equation}
and
\begin{equation}
H_Z(Z_m,T)=-\frac{1}{2}(H_{ZZZ}(Z_m,T))(Z_m-1)^2+H_{ZZ}(Z_m,T)(Z_m-1).
\end{equation}
In doing so, the solution of (\ref{evolutionND}) for the deflection of the sheets in the liquid-filled region automatically satisfies the boundary conditions at $Z=1$, and prescribes the deflection of the sheets in the liquid-free region $Z_m < Z \le 1$. The boundary conditions are summarized in Table~\ref{problemsummary}.
Finally, we take for the initial conditions $H(Z,0)=1$ and $Z_m(0)=0.001$.

\begin{table}
\centering
\caption{Boundary conditions for (\ref{evolutionND}) in each of the three temporal regimes: Regime I (open ends), Regime II (ends in contact), and Regime III (ends coalescing).}
\begin{tabular}{|c|c|c|}
 \hline 
 Regime I & Regime II & Regime III\\
 \hline \hline
  \multicolumn{3}{|c|}{$H_{ZZ}(0,T)=0$}\\ \hline
    \multicolumn{3}{|c|}{$H_{ZZZZ}(0,T)=0$}\\ \hline
    $H_{ZZZ}(0,T)=0$ & & \\ \hline
   & $H(0,T)=0$& $H(Z\le Z_c,T)=0$ \\ \hline
    & & $H_Z(Z\le Z_c,T)=0$ \\ \hline
    \multicolumn{3}{|c|}{$H_{ZZZZ}(Z_m,T)=-\mathscr{E}\cos \theta_d/H(Z_m)$}\\ \hline
    \multicolumn{3}{|c|}{$H(Z_m,T)=1-\frac{1}{3}(H_{ZZZ}(Z_m,T))(Z_m-1)^3+\frac{1}{2}H_{ZZ}(Z_m,T)(Z_m-1)^2$}\\ \hline
    \multicolumn{3}{|c|}{$H_Z(Z_m,T)=-\frac{1}{2}(H_{ZZZ}(Z_m,T))(Z_m-1)^2+H_{ZZ}(Z_m,T)(Z_m-1)$}\\ \hline
\end{tabular}
\label{problemsummary}
\end{table}%

\section{Solution of the free-boundary problem}\label{sec:solution}
The solution to the free-boundary problem defined in \S \ref{sec:theory} is divided into two stages. 
For early times ($T\ll 1$) the deflection of the sheets is small, time enters only through the boundary conditions at $Z_m(T)$, and the free-boundary problem admits the approximate solution
\begin{equation}
H(Z,T)= 1-\frac{\mathscr{E} Z^5}{120Z_m}+\left(\frac{Z_m^3}{8}-\frac{Z_m^2}{3}+\frac{ Z_m}{4}\right)\mathscr{E} Z
+\left(\frac{Z_m}{6}-\frac{Z_m^3}{30}-\frac{1}{6}\right)\mathscr{E} Z_m, \label{earlyprofile}
\end{equation} 
for the deflection of the sheets in the liquid-filled region, and
the liquid advances as though it were between rigid sheets. Hence, with $\hat{Z}_m = Z_m \mathcal{B}=z_m/\ell_{cg}$ and $\hat{T}=T \mathscr{E}\mathcal{B}^2=t/\tau_r$, we find
\begin{equation}
\frac{d\hat{Z}_m}{d\hat{T}}=\frac{\cos \theta_d}{\hat{Z}_m}-1, \label{speedandposition}
\end{equation}
in accordance with classical capillary rise. When $\cos \theta_d=1$, (\ref{speedandposition}) can be integrated to give an implicit expression for $\hat{Z}_m(\hat{T})$:
\begin{equation}
\hat{T}=-\hat{Z}_m - \ln(1-\hat{Z}_m). \label{classicalrise}
\end{equation}
At early times, (\ref{classicalrise}) reduces to $\hat{Z}_m(\hat{T})=\sqrt{2\hat{T}}$, or equivalently, $Z_m(T)=\sqrt{2\mathscr{E} T}$ for $Z_m \ll \mathcal{B}^{-1}$. When $\cos \theta_d\neq1$, the meniscus is slowed down at early time and reaches the classical behaviour at late times.

For late times, when the deflection of the sheets is significant, we solve numerically the free-boundary problem using a method similar to the one described in \cite{Ari10}. We transform the time-dependent domain $0\le Z \le Z_m(T)$ to the fixed domain $0\le S\le 1$ via the substitution $S = Z/Z_m(T)$, and rewrite the time and spatial derivatives in (\ref{velocityND}) and (\ref{evolutionND}) accordingly. Then, the resulting equations and boundary conditions are discretized using an implicit finite-difference scheme and solved numerically in MATLAB\textsuperscript{\textregistered}. We use for initial conditions the profile predicted by (\ref{earlyprofile}) and the meniscus speed given by (\ref{speedandposition}).

\begin{figure}
\begin{center}
\includegraphics[width=0.8\textwidth]{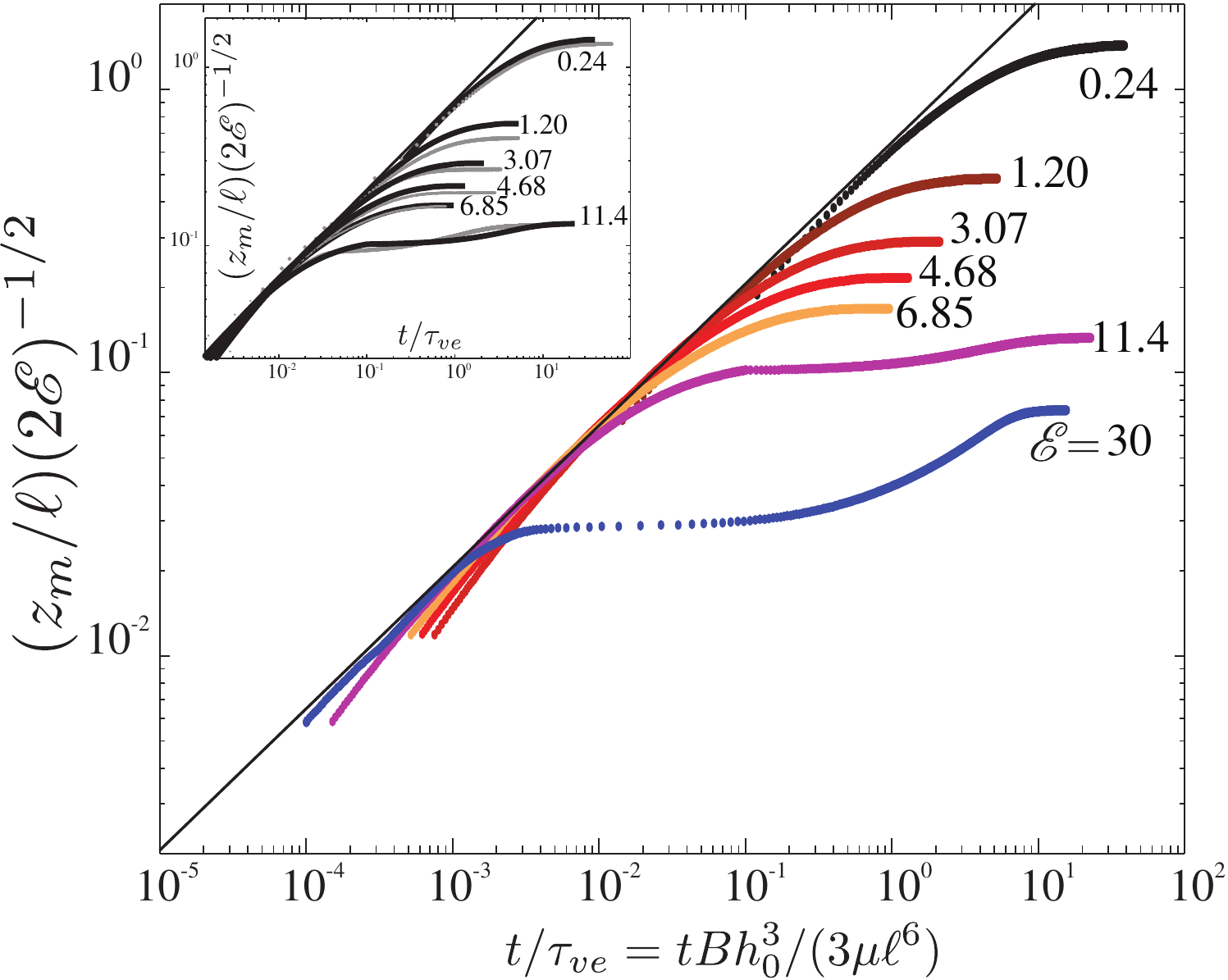}
\caption{Rescaled position of the meniscus $(z_m/\ell) (2\mathscr{E})^{-1/2}$ as a function of time $t/\tau_{ve}=t\frac{Bh_0^3}{3\mu\ell^6}$ for numerics. The solid line corresponds to $Z_m=\alpha \sqrt{2\mathscr{E}T}$ with a pre-factor of $\alpha\simeq 0.65$ due to the dynamic contact angle effect given by (\ref{tanner}). Inset: direct comparison between experimental data (grey points) and numerical solutions (black lines).}
\label{fig.5b}
\end{center}
\end{figure}

The evolution of the position of the meniscus obtained from our numerical simulation is shown in Fig.~\ref{fig.5b}. Comparison with experiments (Fig.~\ref{fig.5a}) shows that our model captures the essential features of the experiments, the collapse onto $\sqrt{\mathscr{E}T}$ at early times, and the observed plateau and subsequent acceleration at late times and for large $\mathscr{E}$. At early times, the position of meniscus follows the expected diffusive-like behaviour $Z_m=\alpha \sqrt{2\mathscr{E}T}$. The pre-factor $\alpha=0.65<1$ arises solely from the deceleration induced by the dynamic contact angle (\ref{tanner}) and is identical in the experiments and in the numerical solutions. Direct comparison between the experimental and numerical results (inset in Fig.~\ref{fig.5b}) shows good quantitative agreement between the experimental data and the corresponding numerical solutions. The discrepancy at late times may result from the difficulty in accurately measuring the bending stiffness $B$ or the presence, in Regimes II and III, of a thin liquid layer between the sheets, which is neglected in the numerics and effectively increases the gap in the experiments. We also verify the scaling law for the cross-over time $T'$ (inset in Fig.~\ref{fig.5a}) with a numerical pre-factor of 0.25 corresponding to a cross-over when the meniscus reaches $z_m\simeq 0.5~ \ell_{ec}$. For $\mathscr{E}>35$, the equilibrium shape is not achieved in the simulations, owing to the meniscus speed reaching nearly zero in the plateau region.

\section{Discussion}\label{sec:discussion}
\begin{figure}
\begin{center}
\includegraphics[width=0.8\textwidth]{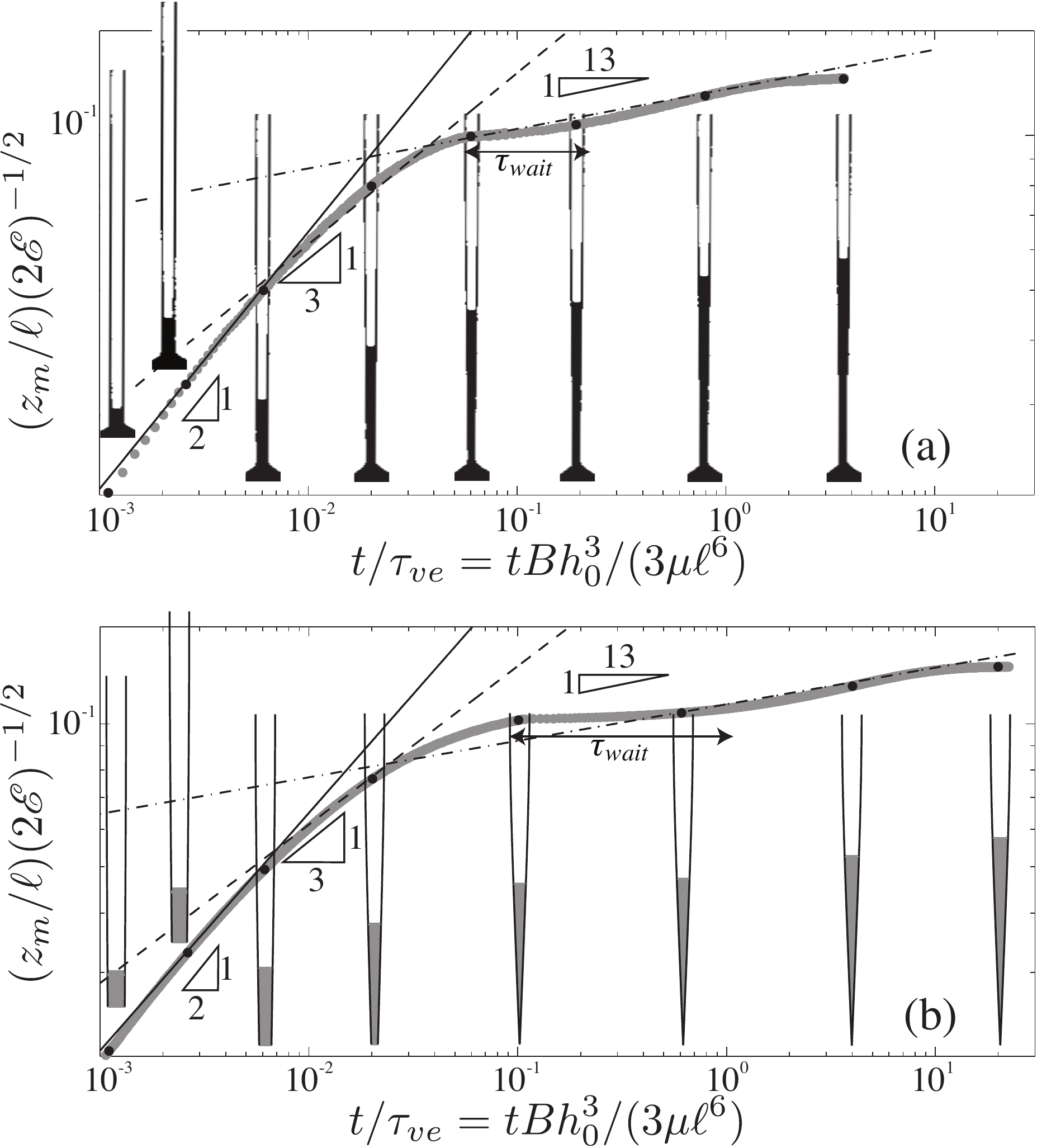}
\caption{Position of the meniscus $(z_m/\ell) (2\mathscr{E})^{-1/2}$ versus time for $\mathscr{E}=11.4$, $\mathcal{B}=2$ for (a) experiments and (b) numerics. The solid lines correspond to $Z_m=0.65\sqrt{2\mathscr{E}}~T^{1/2}$. When $T=T'$, the meniscus follows the power law expected in a wedge geometry $Z_m=0.28\sqrt{2\mathscr{E}}~T^{1/3}$ (dashed lines). At later times, the channel formed by the two sheets adopts a complex geometry and the meniscus roughly follows a power law $Z_m\propto T^{1/13}$ (dash-dotted lines) with a pre-factor of 0.12 (experiments) and 0.11 (numerics). Insets show (a) processed experimental photographs and (b) computed profiles. The corresponding times are denoted by the symbols ($\bullet$) on the grey curves.}
\label{fig.6}
\end{center}
\end{figure}

The interpretation of our results may be made by considering the geometry of the channel. As the sheets are deflected, the gap evolves from being uniform to having a diverging geometry when the sheets touch. The asymptotic behaviours of the position of the meniscus are shown in Fig.~\ref{fig.6}a (experiment) and Fig.~\ref{fig.6}b (numerics) for matching values of $\mathscr{E}$ and $\mathcal{B}$.
We first consider the experimentally observed dynamics (Fig. \ref{fig.6}(a)). At early times, the meniscus obeys the diffusive law $Z_m\propto T^{1/2}$ expected for a uniform gap in the absence of gravity. When $T=T'$, the deflection of the sheets becomes important and the gap resembles a wedge. The meniscus follows the power law expected in a wedge geometry $Z_m\propto T^{1/3}$ \cite[]{War04}. We then observe a plateau of duration $\tau_{\mathrm{wait}}$ during which the meniscus advances slowly. The channel formed by the two sheets adopts a complex geometry, and the data roughly follows a power law $Z_m\propto T^{1/13}$. This behaviour is consistent with imbibition in a diverging channel having shape $h(z)=h_0+ \beta z^n$, where the time-dependence of the meniscus position is given by $Z_m\propto T^{1/(2n+1)}$ \cite[]{Rey08}. Our results are close to $n=6$, although any value of $n>1$, hence an exponent of $1/(2n+1)<1/3$, is permissible. We note, however, that gravity will slow down the rise of liquid and so the analogy is qualitative at best.

The asymptotic behaviours found experimentally are recovered in the numerical results (Fig.~\ref{fig.6}(b)), which allow us to accurately investigate the changes in the shape of the sheets during imbibition. In particular, we identify the time at which the sheets touch ($T=1.18\cdot 10^{-1}$) as the time at which the meniscus reaches a plateau, before accelerating owing to the new geometry of the channel. We obtain good agreement between the experimental data and the corresponding numerical solution. Nevertheless, the duration of the plateau $\tau_{\mathrm{wait}}$ is overestimated in the numerical version of the imbibition, due to the sensitivity of the numerical scheme as the speed of the meniscus approaches zero.


In closing, we have provided a framework to study the dynamics of capillary-driven flow between flexible boundaries, with particular attention given to a model system where capillarity, elasticity, and gravity compete. We have identified the relevant length and time scales, and shown that the time to reach equilibrium sharply increases with increasing elasticity of the boundaries.
Since so many natural phenomena involve elastocapillary effects, it is reasonable to think that the dynamical features we describe here will be useful for rationalizing the time-dependent behaviour of a variety of biological and industrial systems. 



\begin{acknowledgments}
C.D. and H.A.S. gratefully acknowledge financial support from Unilever Research; J.M.A. acknowledges the National Science Foundation Mathematical Sciences Postdoctoral Research Fellowship Program. We thank D. Vella, and P. Warren and A. Lips at Unilever Research for helpful conversations.
\end{acknowledgments}

\bibliographystyle{jfm}
\bibliography{elastocaprise_paper_revision8}

\end{document}